\documentclass[twocolumn,pre,superscriptaddress]{revtex4}
\usepackage{tikz}
\usepackage{graphicx}
\usepackage{color} 
\usepackage{transparent}
\usepackage{psfrag}
\usepackage{dcolumn}
\usepackage{amsmath}
\usepackage{amssymb}
\usepackage{latexsym}
\usepackage{hyperref} 
\usepackage{booktabs}
\usepackage{latexsym}
\usepackage{xcolor}
\begin{document}

% define short cuts !!
\newcommand{\tikzcircle}[2][red,fill=red]{\tikz[baseline=-0.5ex]\draw[#1,radius=#2] (0,0) circle ;}%
\def\bea{\begin{eqnarray}}
\def\eea{\end{eqnarray}}
\def\beq{\begin{equation}}
\def\eeq{\end{equation}}
\def\f{\frac}
\def\k{\kappa}
\def\e{\epsilon}
\def\ve{\varepsilon}
\def\be{\beta}
\def\D{\Delta}
\def\h{\theta}
\def\t{\tau}
\def\a{\alpha}

\def\cDa{{\cal D}[X]}
\def\cD{{\cal D}[x]}
\def\cL{{\cal L}}
\def\cLo{{\cal L}_0}
\def\cLa{{\cal L}_1}

\def\Re{{\rm Re}}
\def\sj{\sum_{j=1}^2}
\def\rk{\rho^{ (k) }}
\def\rek{\rho^{ (1) }}
\def\cek{C^{ (1) }}
\def\rz{\rho^{ (0) }}
\def\rt{\rho^{ (2) }}
\def\rtb{\bar \rho^{ (2) }}
\def\trk{\tilde\rho^{ (k) }}
\def\trek{\tilde\rho^{ (1) }}
\def\trz{\tilde\rho^{ (0) }}
\def\trt{\tilde\rho^{ (2) }}
\def\r{\rho}
\def\tD{\tilde {D}}

\def\s{\sigma}
\def\kb{k_B}
\def\bF{\bar{\cal F}}
\def\F{{\cal F}}
\def\la{\langle}
\def\ra{\rangle}
\def\nn{\nonumber}
\def\up{\uparrow}
\def\dn{\downarrow}
\def\S{\Sigma}
\def\dg{\dagger}
\def\d{\delta}
\def\p{\partial}
\def\l{\lambda}
\def\L{\Lambda}
\def\G{\Gamma}
\def\o{\Omega}
\def\w{\omega}
\def\g{\gamma}

\def\jv{ {\bf j}}
\def\jr{ {\bf j}_r}
\def\jd{ {\bf j}_d}
\def\jdd{ { j}_d}
\def\noi{\noindent}
\def\a{\alpha}
\def\d{\delta}
\def\p{\partial} 
\def\hf{\frac{1}{2}}

\def\la{\langle}
\def\ra{\rangle}
\def\e{\epsilon}
\def\n{\eta}
\def\g{\gamma}
\def\break#1{\pagebreak \vspace*{#1}}
\def\hf{\frac{1}{2}}

\def\rv{{\bf r}}
\def\pc{\phi_c}
\def\rb{\bar\rho}
%% Definitions done-----------------------

\title{Cross-linker mediated compaction and local morphologies in a model chromosome}
%\title{Cross-linker mediated compaction, loop formation and zippering in a model chromosome}
%
\author{Amit Kumar}
\email{amit@iopb.res.in}
\affiliation{Institute of Physics, Sachivalaya Marg, Bhubaneswar 751005, India
}
\affiliation{Homi Bhaba National Institute, Anushaktigar, Mumbai 400094, India}
\author{Debasish Chaudhuri}
\email{debc@iopb.res.in}
\affiliation{Institute of Physics, Sachivalaya Marg, Bhubaneswar 751005, India
}
\affiliation{Homi Bhaba National Institute, Anushaktigar, Mumbai 400094, India}

\date{\today}

\begin{abstract}
Chromatin and associated proteins constitute the highly folded structure of chromosomes. We consider a self-avoiding polymer model of the chromatin, segments of which may get cross-linked via protein binders that repel each other. The binders cluster together via the polymer mediated attraction, in turn, folding the polymer.  Using molecular dynamics simulations, and a mean field description,  we explicitly demonstrate the continuous nature of the folding transition, characterized by unimodal distributions of the polymer size across the transition. At the transition point the chromatin size and cross-linker clusters display large fluctuations, and a maximum in their negative cross-correlation, apart from a critical slowing down. Along the transition, we distinguish the local chain morphologies in terms of topological loops, inter-loop gaps, and zippering.  The topologies are dominated by simply connected loops at the criticality, and by zippering in the folded phase. 

\end{abstract}

\maketitle

\section{Introduction}

Chromosomes consist of DNA and associated proteins. Long DNA chains with lengths that vary from millimeters (bacteria) to meters (mammals) are compacted and organized within micron sized bacterial cells or cell nuclei of eukaryotes. The DNA must be compacted by several orders of magnitude, and still allow information processing in terms of gene expression in both pro- and eukaryotes~\cite{Alberts2015, Wolffe2012}, and also replication in bacteria~\cite{Thanbichler2005}. DNA associated proteins play a crucial role in such processes~\cite{Alberts2015, Wang2011, Wolffe2012,Thanbichler2005}. In the smallest scale, DNA double helix wraps around histone octamers forming a bead on string chromatin structure with a connected set of nucleosomes~\cite{Luger1997}. In bacteria, histone like nucleoid structuring (H-NS) protein dimers bind to DNA~\cite{Dame2006, Wiggins2009}. It may be noted that the positive charges on most DNA binding proteins provide non-specific affinity to negatively charged DNA~\cite{Alberts2015,Wolffe2012}.  

Higher order structure formation involves bringing together of contour-wise distant parts of the chromatin into spatial contact to form loops~\cite{Worcel1972, Holmes2000, Zimmerman2006, VanDerValk2014, Rao2014, Dame2016}. This is observed in all domains of life, in bacteria~\cite{Luijsterburg2006a, Dame2005a}, archea~\cite{Peeters2015a} and eukaryotes~\cite{Rao2014, Bickmore2013,Nasmyth2009,Phillips2009}.  % 
Such loops may be maintained by proteins cross-linking spatially proximal chromatin segments~\cite{Annunziatella2018, LeTreut2016, Brackley2017, Brackley2013, Johnson2015, Brackley2016, Brackley2017, Marenduzzo2007, Kolomeisky2015, Barbieri2012, Nicodemi2009, Nicodemi2014, Fudenberg2012}, or by extrusion~\cite{Fudenberg2016,Sanborn2015,Alipour2012, Brackley2017c}. A number of proteins are identified that stabilize these loops into separate topological domains~\cite{Brocken2018,Goloborodko2016,Song2015}. In eukaryotes cohesin and CTCF are identified as chromatin loop regulators~\cite{Fudenberg2016,Rao2014, Nasmyth2009,Phillips2009}.  In bacteria, loops are stabilized in part by non-specific cross-linkers such as H-NS, Lrp and SMC proteins~\cite{Dame2005, Dame2016, Dame2018}. First direct evidence of the chromosomal loops were found in electron microscopy experiments~\cite{Delius1974, Kavenoff1976, Trun1998, Postow2004}. Complementary experiments using chromosome conformation capture techniques provide contact maps that exhibit spatial contacts between different genes along the DNA contour~\cite{Lieberman-Aiden2009, Le2013, Marbouty2014}. The biological function of chromosomes are often related to their local morphology~\cite{Thanbichler2005, Brocken2018},  as spatially proximal genes, irrespective of their location along the DNA contour, can be regulated together~\cite{Cremer2001a, Dekker2008a, Dillon2010, Meyer2018, Fritsche2012, Llopis2010b}. Given their structural complexity, morphologies of long folded chains are often analyzed in terms of their generic topological features~\cite{Mizuguchi1995, Bailor2010, Cavalli2013, Mashaghi2014}.

During interphase chromosomes display several universal properties, e.g., scale free nature in subchain extension~\cite{Munkel1999}, and average contact probabilities~\cite{Lieberman-Aiden2009} similar to  homopolymers but with exponents that differ from simple chains~\cite{Gennes1979, Rubinstein2003}.  
At small separations the chromosomal architecture is determined by its bending rigidity, while at long range they show behavior typical of fractal globules~\cite{Jost2017, Rosa2008, Mirny2011, Lieberman-Aiden2009, Grosberg1988, Lesage2018, Victor1994}. The contact maps reveal topologically associated domains at smaller genomic separations ($ \lesssim 1$\,Mbp)~\cite{Rao2014,Dixon2012}, and cell type specific checker-board patterns at larger scales~\cite{Lieberman-Aiden2009}. The relationship between chromatin structure and function have been considered explicitly using heteropolymer models to understand the sequence specific aspects of chromosomal organization~\cite{Barbieri2012, Brackley2016, Ganai2014, Jost2014, Olarte-Plata2016, Tark-Dame2014, Zhang2015, Fudenberg2016, Chiariello2016, Agarwal2017, Agarwal2018, Agarwal2018a, Agarwal2018b, Agrawal2018, Ramakrishnan2015}. 

Modeling chromatin as a homopolymer, generic effects of its further association with proteins have been studied using effective attraction between its segments~\cite{Tark-Dame2012, Scolari2015, Scolari2018}, or  through explicit consideration of its interaction with diffusing proteins~\cite{LeTreut2016, Barbieri2012, Brackley2013, Johnson2015, Brackley2016, Brackley2017}. The interaction eventually leads to folding of the chromatin~\cite{Nicodemi2009,Nicodemi2014,Barbieri2012,Lesage2018}. 
In this context, the classic polymer physics problem of coil-globule transition has received renewed interest. The Flory-Huggins theory of coil-globule transition suggests coexistence lines of coil-rich and globule-rich phases ending at a critical point with changing solubility~\cite{Gennes1979}. 
Extensions of Flory-like theory predicted first order or a continuous coil-globule transition depending on parameter values~\cite{Ptitsyn1965, Eisner1969, DeGennes1975}.  Other theoretical approaches predicted similar behavior~\cite{Lifshitz1978,Lifshitz1976}. 
While explicit consideration of binders as attractive co-solvent suggested a continuous coil-globule transition for flexible chains in numerical simulations, analysis of the same within Flory-Huggins theory predicted a first order transition~\cite{LeTreut2016}. A recent mean field approach that incorporates fluctuations of co-solvent density showed that  the nature of the coil-globule transition with the increase in co-solvent density depends on the the kind of polymer- co-solvent interaction. When the interaction strength between the polymer  and co-solvent is purely repulsive the predicted transition is continuous, whereas it turns out to be a first order transition if the interaction is purely attractive~\cite{Budkov2014}.

In this paper, we model the chromatin as a self-avoiding chain and explicitly consider its non-specific attraction with diffusing binder proteins that repel each other. 
We perform molecular dynamics simulations in the presence of a Langevin heat bath to fully characterize the binder mediated folding transition, and associated local topologies of the chromosome.  
The binders cross-link different segments of the chromatin, bringing them together. As a result more binders accumulate, forming clusters. Such clusters, in turn, fold the polymer. Using numerical simulations and a mean field description we show that the folding is a continuous transition, mediated by a linear instability towards formation of large clusters of cross-linkers. While the linear stability prediction shows reasonable agreement with simulations for growth of cluster size before the transition, the mean field prediction for chromosome size display better agreement after the transition, as fluctuations get suppressed in the globule. The polymer size distribution shows a single maximum across the transition, signifying that there is no metastable phase on the other side of the transition. 
The criticality is characterized by large and slow fluctuations -- the fluctuation amplitude and  time-scale increase with system size. At criticality, the cross-correlation between chromosome size and cross-linker density shows a negative maximum. 

We further analyze topologies of the chromatin loops that are formed due to binder cross-linking, identifying the simply connected and higher order loops. The average number of simply connected loops show a maximum at the critical point, while the relative probabilities of loops of different orders change qualitatively with increasing cross-linker density. The first order loop-sizes show power law distributions with exponents that change monotonically across the transition. The gaps between such loops, in contrast, follow exponential distributions. The mean gap size hits a minimum at the critical point. Apart from forming loops, the binders may also zipper contiguous segments belonging to different parts of the chromatin. As we show, the mean number of zippering displays a sigmoidal behavior along the coil-globule transition, with saturation at large binder densities.

In Sec.~\ref{model}, we present the model and details of numerical simulations. 
In Sec.\ref{results} we discuss simulation results identifying the coil-globule phase transition in the model chromatin chain, and clustering of polymer-bound cross-linkers. The transition and clustering are interpreted in terms of a mean field description and linear stability analysis. Finally, in Sec.~\ref{sec:morpho} we characterize the local morphology of chromosomes in terms of contact probability, loop topologies of various orders, and zippering.  We conclude presenting a connection of our results to experimentally verifiable predictions in Sec.~\ref{sec:concl}.

%%%%%%
\begin{figure}[t]
\begin{center}
\includegraphics[width=8cm]{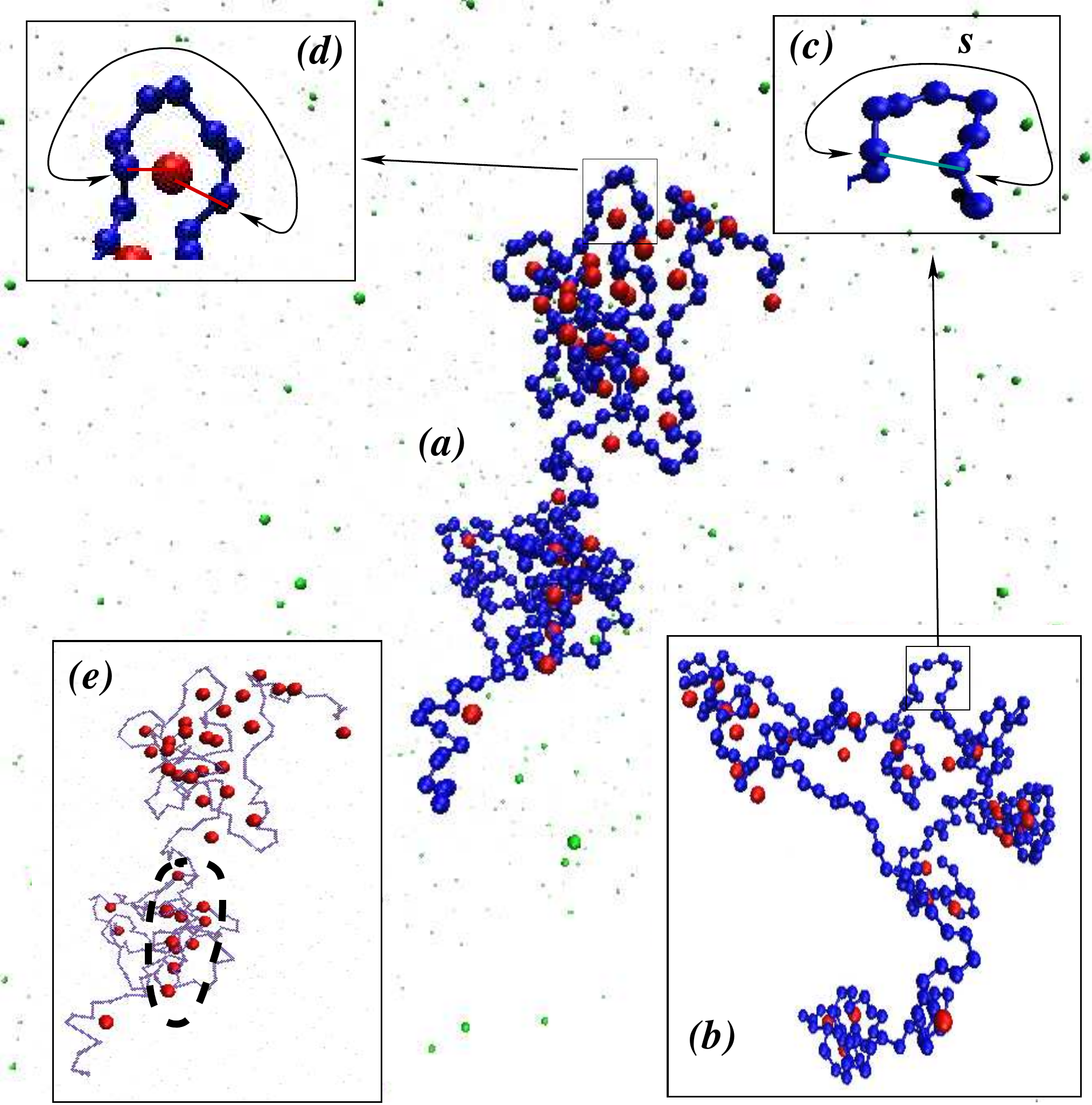} 
\caption{(color online)~Representative snapshots of the model chromosome with $N=256$ bead chain and binders at the transition point $\phi_c=1.57 \times 10^{-3}$, where large conformational fluctuations are observed.  
($a$)~A relatively compact conformation. The chromatin is shown by blue monomers connected by bonds. 
The cross-linkers attached to the chromatin are shown as red beads, while the freely diffusing binders are shown as green beads. 
($b$)~One relatively open conformation. 
($c$)~A magnified portion of ($b$) shows a contact formation denoted by the aqua-green bar. A monomer pair, contour-wise separated by $s$, have come within $r_c$ forming the contact. 
($d$)~A magnified portion of ($a$) shows loop formation by a  polymer bound cross-linker (red  bead). The red bars indicate the bonds that it forms. 
The line with arrow-heads identifies a simply connected loop~(for further details see Sec.~\ref{sec:loop}). 
($e$)~Clusters of polymer bound cross-linkers in ($a$). For better visibility of cross-linkers, the chromatin is represented by a faded line. The thick dashed circle identifies one cluster. % of cross-linkers. 
}
\label{fig:snapshot}
\end{center}
\end{figure}

%%%%%%

%%%%%%%%%%%%
\begin{figure*}[t]
\begin{center}
\includegraphics[width=7cm]{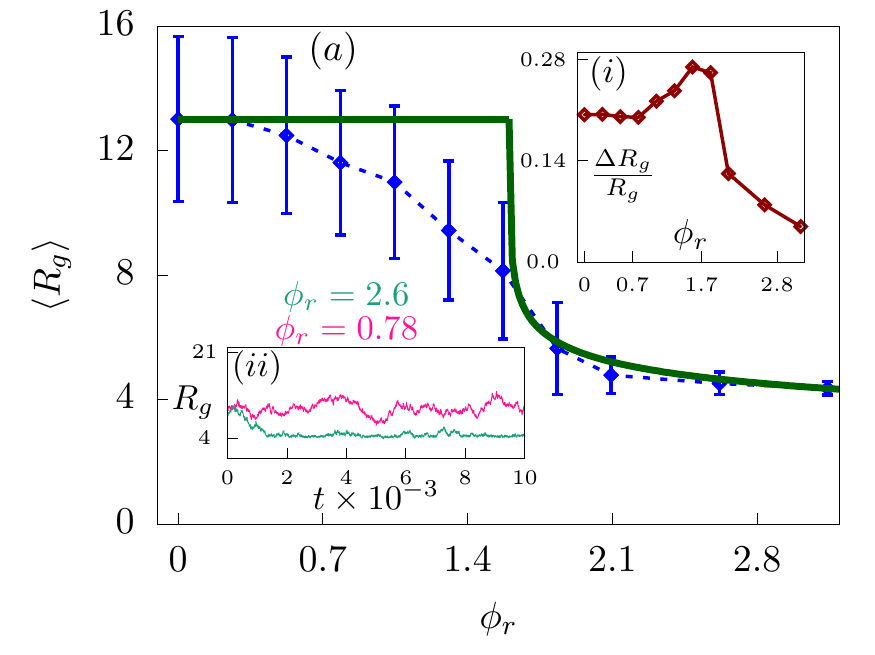}
\includegraphics[width=7cm]{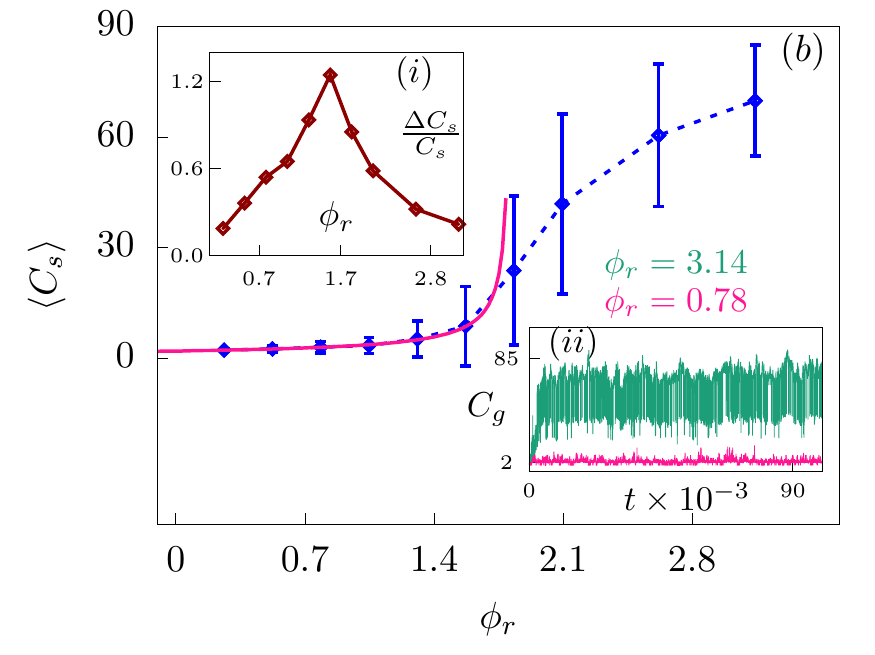}
\includegraphics[width=3.5cm]{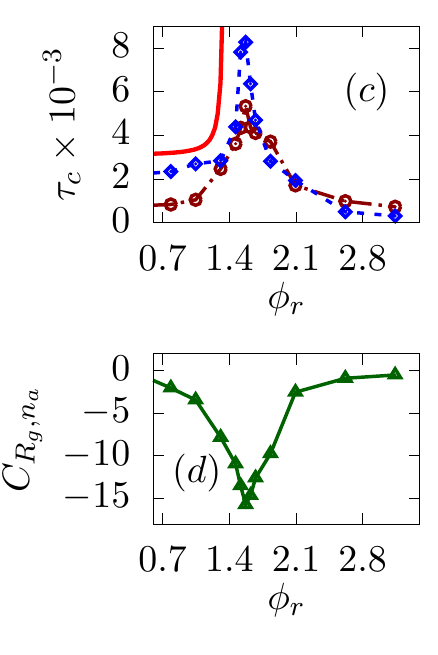}
\caption{(color online) The coil-globule transition as a function of ambient cross-linker density $\phi_c$, expressed in terms of $\phi_r = \phi_c \times 10^{3}$. 
($a$)\,The decrease in mean radius of gyration of polymer $\la R_g \ra$ with $\phi_c$, the data are shown by $\diamond$ and error bars, captures a coil-globule transition. The blue dashed line is a guide to eye. 
The mean field prediction $\la R_g \ra=$ constant $= \la R_g\ra (\phi_c=0)$ at $\phi < \phi_c^\ast=1.57\times 10^{-3}$. At higher densities, 
simulation results for $\la R_g \ra$ show reasonable agreement with Eq.\ref{eq:rg} with fitting parameter ${u/v}=0.1$.
The two curves are shown by  green solid lines.
At the transition point $\phi_c^\ast$,  relative fluctuation of polymer size  $\Delta R_{g}/R_{g}$  
shows a maximum (inset ($i$)). 
The equilibration of $R_g$ with time $t$ at two densities $\phi_c=0.78\times 10^{-3}$~(pink),$\, 2.6 \times 10^{-3}$~(green) are shown in inset ($ii$). 
($b$)\,The mean size of the polymer bound clusters of cross-linkers $\la C_s \ra$  increases with $\phi_c$. The data are shown by $\diamond$ and error bars. The blue dashed line is a guide to eye. 
At $\phi_c < \phi_c^\ast$, the data show reasonable agreement with $\la C_s \ra= {\cal A} [(1-\phi_c/\phi_c^\ast ]^{-3/4}$ , using  ${\cal A} = 1.9$.
The relative cluster size fluctuation $\Delta C_{s}/C_{s}$ shows a sharp maximum at the transition point $\phi_c^\ast$ (inset ($i$)). 
Inset ($ii$) shows how the instantaneous mean cluster size $C_g$ equilibrates with time $t$ at two cross-linker densities $\phi_c=0.78\times 10^{-3}$~(pink)$,\, \pi \times 10^{-3}$~(green). 
($c$)~Correlation times $\t_c = \t_{R_g}\,(\diamond), \t_{n_a}\,(\circ)$ are obtained from auto-correlation functions of polymer radius of gyration $R_g$, and the total number of chromatin-bound cross-linkers $n_a$. They reach their maximum values at the transition point $\phi_c^\ast$. The solid (red) line is a shifted plot of the scaling form $(1-\phi_c/\phi_c^\ast)^{-3/2}$ added to a constant background, with the shift aimed at better visibility.
($d$)~The negative values of the cross-correlation coefficient $C_{R_g,n_a}$ between fluctuations in $R_g$ and $n_a$ show anti-correlation, with the amplitude maximizing at the transition point $\phi_c^\ast$.  
}
\label{fig_meanRg}
\end{center}
\end{figure*}
%%%%%%%%%%%%

\section{Model}
\label{model}
We use a self-avoiding flexible chain model of the chromatin. The bead size is assumed to be larger than the Kuhn length.  The chain connectivity is maintained by finitely extensible nonlinear elastic (FENE) bonds between consecutive beads,  $ { U_{\rm FENE}(r_{i+1,i}) }= -\frac{k}{2} R^{2} \ln[1-(r_{i+1,i}/R)^{2}]$  where $k$ and $R$ fix the bond, and $r_{ij} = |\rv_i - \rv_j|$ denotes separation between $i$-th and $j$-th bead. The self avoidance is implemented via the Weeks-Chandler-Anderson potential, ${ U_{\rm WCA} (r_{ij}) }= 4{\epsilon}[(\sigma/r_{ij})^{12}-(\sigma/r_{ij})^{6}+0.25]$ between beads separated by distance $r_{ij}  < 2^{1/6}\s $, else ${ U_{\rm WCA} (r_{ij})} = 0$~\cite{Weeks1971a}. Thus {  $ U =  U_{\rm FENE} +  U_{\rm WCA} $} defines the polymer~\cite{Grest1986a}.  
The repulsion between cross-linkers are modeled through the same $U_{WCA}$ interaction. 
The energy and length scales are set by {\color{blue} $\e$} 
and $\s$ respectively. 
The FENE potential is set by { $k=30.0\, \e/\s^2$, $R=1.6\, \s$}.  

The interaction between cross-linkers and monomers is modeled through a truncated and shifted Lennard-Jones potential,  $U_{\rm shift} (r) = \be U_{\rm LJ} (r) - U_{\rm LJ} (r_c) $  for $r<r_c$ and $U_{\rm shift} (r) =0$ otherwise, where $U_{\rm LJ} (r) = 4{\epsilon_m}[(\sigma/r)^{12}-(\sigma/r)^{6}]$, with  {$\e_m = 3.5\, \e$}, and $r_c = 1.5\,\s$. 
The choice of $\e_m$ is stronger than the typical hydrogen bonds~($1.2 \, \kb T$) and provides better stability~\cite{Watson2003}, e.g., as for transcription factors~\cite{Barbieri2012}, however, allows equilibration through attachment- detachment kinematics over the simulation time scales. 
The bond between a cross-linker and a monomer is formed if they come within the range of attraction $r_c$. 
A single cross-linker may bind to multiple monomers, capturing the presence of multiple DNA binding domains in a number of regulatory proteins~\cite{Barbieri2012}.

{%\color{blue} 
The molecular dynamics simulations are performed using  the standard velocity-Verlet algorithm~\cite{Frenkel2002} using time step $\d t = 0.01 \t$, where $\t = \s \sqrt{m/\e}$ is the characteristic time scale. The mass of the particles are chosen to be $m=1$. The temperature of the system is kept constant at $T = 1.0\, \e/\kb$ by using a Langevin thermostat~\cite{Grest1986a} characterized by an isotropic friction constant $\g = 1/\t$, as implemented by the ESPResSo molecular dynamics package~\cite{Limbach2006}. Similar methods have been successfully used earlier in simulation of polymers in various contexts~\cite{Duenweg1995}.   Note that the diffusion of a single bead over its size $\s$ takes a time $\g \s^2/\kb T$, which is the same as the characteristic time $\t$.}

Unless stated otherwise, in this paper, we consider a $N=256$ bead chain. 
Its typical size in absence of binders is given by the radius of gyration $R^0_g=(13.02 \pm 2.65)\,\s$. 
The largest fluctuations in its end to end separation are restricted within $80\, \s$ (data not shown). To avoid any possible boundary effect, we perform simulations  in a cubic volume of significantly bigger size with sides of $L=200\,\s$,  and implement periodic boundary condition.   We vary the total number of cross-linkers from $N_c= 0$ to $6000$ that changes the dimensionless cross-linker density from $\phi_c = \frac{4}{3}\pi \sigma^3 N_c/L^3 =0$ to $\pi \times 10^{-3}$. 
The approach to equilibrium is followed over $10^6\,\t$, longer than the longest time taken for equilibration near the transition point. The  analyses 
are performed over further runs of $10^6\,-10^7\, \t$.
A couple of representative equilibrium configurations are shown using VMD~\cite{HUMP96} in Fig.\ref{fig:snapshot} illustrating polymer contacts, loop formation, and clustering of cross-linkers.
The system size dependence is studied using a restricted set of simulations, as simulating longer chains requires longer equilibration, larger simulation box and larger number of cross-linkers, increasing the simulation time significantly.

\begin{figure*}[t]
\begin{center}
\includegraphics[width=8cm]{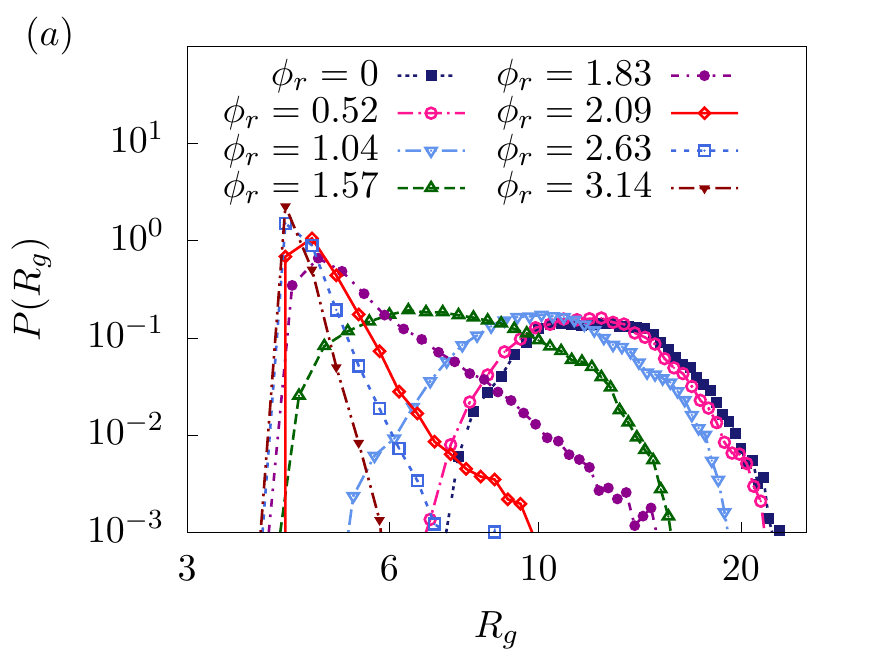} 
\includegraphics[width=8cm]{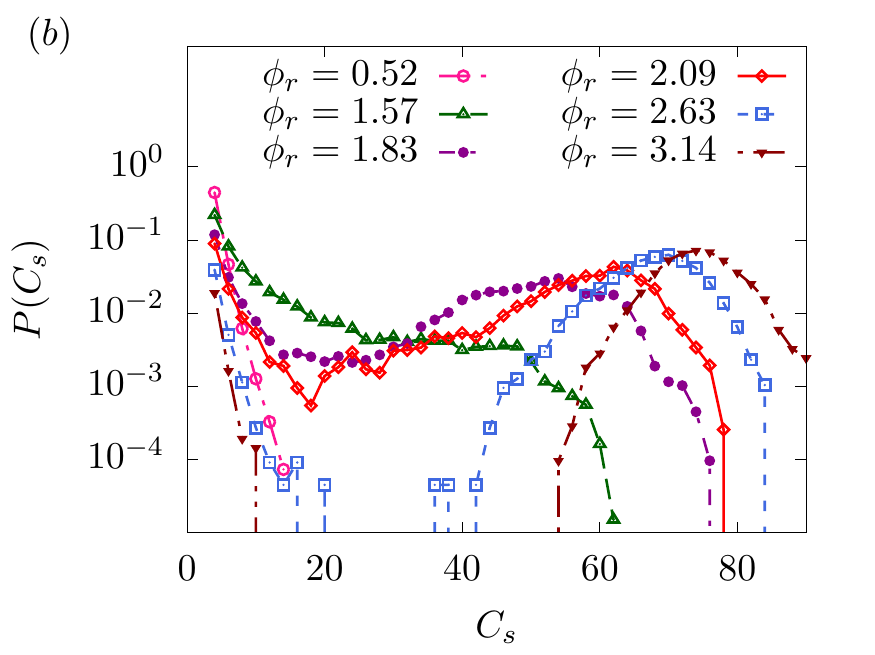}
\caption{(color online)~Probability distributions of the polymer radius of gyration, $P(R_g)$, are plotted at different cross-linker densities denoted by $\phi_r = \phi_c \times 10^{3}$. ($b$)~Corresponding probability distributions of cluster size of cross-linkers, $P(C_s)$,  show coexistence of small and large clusters at densities $\phi_c \geq 1.83 \times 10^{-3}$. }
\label{fig_prob_distbn_rg_cs}
\end{center}
\end{figure*}

\section{Results}
\label{results}
%%%

\subsection{The coil-globule transition}
The passive binders diffuse in three dimensions and attach to polymer segments following the Boltzmann weight. They are multi-valent, typically cross-linking multiple polymer segments. The probability of number of chromatin segments that a binder can cross-link simultaneously shows a maximum that increases from $4$ to $6$ as the average binder concentration is increased~(see Appendix-\ref{sec:valency}). This range overlaps with the typical multiplicity of binding factors like CTCF and transcription factories~\cite{Barbieri2012}.  

As different polymer segments start attaching to a cross-linker the local density of monomers increases, generating a positive feedback recruiting more cross-linkers and as a result localizing more monomers. Such a potentially runaway process gets stabilized, within our model, due to the inter-binder repulsion that ensures the binder-clusters are spatially extended. These clusters are identified using the clustering algorithm in Ref.~\cite{Allen1989}, and the cluster-size is given by the total number of binders in a cluster. Concomitant with such clustering, the polymer gets folded undergoing a coil-globule transition. 

Fig.~\ref{fig_meanRg}($a$) shows the transition in terms of the decreasing radius of gyration $\la R_g \ra$ of the model chromatin with increase in the average cross-linker density $\phi_c$ in the environment.  
  The solid (green) line shows the mean field prediction that we present in Sec.~\ref{sec:mft}.   
 The transition point, $\phi_c^\ast=1.57 \times 10^{-3}$,  is characterized by a maximum in relative fluctuations of the polymer size $\D R_g/R_g = \sqrt{\la R_g^2 \ra - \la R_g \ra^2}/\la R_g \ra$, shown in the inset ($i$) of  Fig.~\ref{fig_meanRg}($a$). The equilibrations of $R_g$ at two representative binder concentrations $\phi_c$ are illustrated in the inset  ($ii$).
 As we show in Fig.~\ref{fig_phicN} in Appendix-\ref{ap:transition},  the relative fluctuations $\D R_g/R_g$ near phase transition increases with polymer size $N$, suggesting divergence in the thermodynamic limit, a characteristic of continuous phase transitions.  
 In  Fig.\ref{fig:snapshot}, the large fluctuations at the phase transition point are further illustrated with the help of two representative conformations: a relatively compact conformation in Fig.\ref{fig:snapshot}($a$), and and a more open conformation in Fig.\ref{fig:snapshot}($b$).

The coil-globule transition occurs concomitantly with the formation of polymer-bound cross-linker clusters. At a given instant, several disjoined clusters may form along the model chromatin~(see Fig.\ref{fig:snapshot}($e$)\,).  The cluster size $\la C_s \ra$ is the average number of binders constituting the clusters. It grows significantly as $\phi_c$ approaches phase transition from below~(Fig.~\ref{fig_meanRg}($b$)). 
The linear stability estimate of cluster size, as discussed in Sec.~\ref{sec:cluster}, is represented by the (pink) solid line in Fig.~\ref{fig_meanRg}($b$). 
The relative fluctuations in cluster size $\D C_s/ C_s = \sqrt{\la C_s^2 \ra - \la C_s \ra^2}/\la C_s \ra$ show a sharp maximum at the phase transition point $\phi_c^\ast$ (inset ($i$)).  Equilibration of the mean cluster size $C_g$, instantaneous average over all clusters, at two $\phi_c$ values are shown in the inset($ii$).

The dynamical fluctuations at equilibrium are characterized by
the auto-correlation functions $C_{R_g}(t)= \la \d R_g(t) \d R_g(0)\ra/\la \d R_g^2 \ra$ of chromatin size $R_g$,  and $C_{n_a}(t)=\la \d n_a(t) \d n_a(0) \ra/\la \d n_a^2 \ra$ of the total number of chromatin-bound cross-linkers, where $\d R_g(t)$ and $\d n_a(t)$ denote instantaneous deviations of the two quantities from their respective mean values~(see Appendix-\ref{sec:corrln}). For the finite sized chain,
the corresponding correlation times $\t_c=\t_{R_g},\, \t_{n_a}$ show sharp increase at $\phi_c^\ast$~(Fig.~\ref{fig_meanRg}($c$)), reminiscent of the critical slowing down~\cite{Chaikin2012}. As is shown in Appendix-\ref{ap:transition}, the correlation time $\t_{R_g}$ grows with the chain-length as $\t_{R_g} \sim N^{9/4}$ suggesting divergence in the thermodynamic limit. 

The  fluctuations in $n_a$ and $R_g$ are anti-correlated, as a larger number of attached cross-linkers reduces the chromatin size. Thus the cross-correlation coefficient $C_{R_g,n_a} = (1/\t_p)\int^{\t_p} dt \la \d R_g(t) \d n_a(t) \ra < 0$. Remarkably, the amount of anti-correlation maximizes at the critical point $\phi^\ast_c$ signifying a large reduction in polymer size associated with a small increase of attached cross-linkers, and vice versa~(Fig.~\ref{fig_meanRg}($d$)). 
A living cell may utilize this physical property for easy conformational reorganization, useful for providing  access to DNA-tracking enzymes in an otherwise folded chrmosome.

The probability distribution of the radius of gyration $P(R_g)$ shows clear unimodal shape across the transition~(Fig.~\ref{fig_prob_distbn_rg_cs}($a$)\,). 
This clearly displays absence of metastable phase on the other side of the transition, characteristic of the continuous transition. 
Note that this observation is in contrast to the mean field prediction of Ref.~\cite{Budkov2014}, while is in agreement with the numerical simulations in Ref~\cite{LeTreut2016}.

Remarkably, the associated distribution of cross-linker cluster sizes $P(C_s)$, shows clear bimodality in much of the $\phi_c$ range scanned across the transition~(Fig.\ref{fig_prob_distbn_rg_cs}($b$)\,), capturing coexistence of clusters of small and large sizes. However, such clusters have similar densities (data not shown) and do not suggest coexistence of two phases. In fact,  as we show in Sec.\ref{sec:cluster}, the assumption of constant binder density within the clusters provides a reasonable description of the growth of mean cluster size through Eq.(\ref{eq:cs})~(see Fig.\ref{fig_meanRg}($b$)\,).  

\subsection{Mean field description}
\label{sec:mft}
In view of the above phenomenology, we present a mean field model based on two coupled fields, the cross-linker density $\phi_c(\rv)$, and the deviation of monomer density due to cross-linkers $\r(\rv) =  \r_m(\rv) - \r_b $, where $\r_b = \s^3 N/(R^0_g)^3$ with $R^0_g$ denotes the radius of gyration of the open chain in absence of cross-linkers. A fraction of cross-linkers are in polymer bound state, $\phi(\rv)$, and the rest constitutes the detached fraction. 
We adopt a free energy density~\footnote{In the analysis of phase transition, near the transition point, a possible bilinear coupling $-z \r \phi$ is neglected in favor of algebraic simplicity.}
\bea
\be f =   \hf  u\, \left(  1-\f{\phi}{\phi_\ast} \right) \r^2   + \f{v}{4} \r^4 + \f{\k}{2} (\nabla \r)^2 + \hf w \phi^2.
\label{eq:conserve}
\eea
The direct repulsion between polymer segments and between cross-linkers are captured by free energy costs $u\r^2/2$ and $w \phi^2/2$ respectively. The bond formation between two polymeric segments via cross-linker proteins is captured by the three body term $\r \,\phi\, \r$ with strength $-u/2\phi_\ast$. 
The quartic energy cost $v \r^4/4$ is introduced to provide thermodynamic stability. The coefficient $\k$ in the gradient term adds free energy cost to the formation of sharp interfaces in local monomer-density. 
The evolution of coupled fields are represented by~\cite{Chaikin2012},
\bea
\f{\p \r}{\p t} &=& M_\r \nabla^2  \left[ u \left(  1-\f{\phi}{\phi_\ast} \right) \r + v \r^3 - \k \nabla^2 \r \right] \nn\\
\f{\p \phi}{\p t} &=& M_\phi \nabla^2  \left[  -\f{u}{2\phi_\ast} \r^2 + w \phi \right] - r(\phi-\phi_0), 
\label{eq:dyn}
\eea
where, the second term in the right hand side of the second equation accounts for the turnover between the attached and detached fractions of the cross-linkers. Here $r=(r_a +r_d)$, $\phi_0 = \o \phi_c$ with $\o= r_a\, /(r_a + r_d)$. The attachment detachment rates $r_{a,d}$ are determined by the interaction and detailed balance condition.  
The coefficients $M_{\r,\phi}$ denote mobilities of the two conserved fields  $\r$ and $\phi$. A similar approach was used earlier in Ref.~\cite{Brackley2017}.
In the uniform equilibrium state $\phi=\phi_0$, and $\r=\r_0$. Using $\phi_0 = \o \phi_c$ and $\phi_\ast = \o \phi_c^\ast$,  if $\phi_c < \phi_c^\ast$ the solution $\r_0=0$ , else 
\bea
\r_0^2 = \f{u}{v} \f{(\phi_c-\phi_c^\ast)}{\phi_c^\ast}.
\label{eq:rho}
\eea

\subsubsection{Chromosome size}

The mean monomer density $\r_m = \s^3 N/ \la R_g \ra^3 = \r_0 + \r_b$.
As $\phi_c \geq \phi_c^\ast$,  using  Eq.(\ref{eq:rho})
one obtains 
 \bea
\la R_g \ra = {R_g^0}  \left[1+N^{4/5} \left( \f{u}{v} \f{ \phi_c - \phi_c^\ast} {\phi_c^\ast} \right)^{1/2} \right]^{-1/3}.
\label{eq:rg}
\eea  
This shows reasonable agreement with simulation results with fitting parameter $u/v = 0.1$~(Fig.~\ref{fig_meanRg}($a$)),
as fluctuations are suppressed in the globule phase~\cite{Lifshitz1978}. In the limit of $\phi_c \gg \phi_c^\ast$, $\la R_g \ra \approx N^{1/3}\s [(u/v)(\phi_c - \phi_c^\ast)/\phi_c^\ast]^{-1/6}$, i.e., an equilibrium globule with $ \la R_g \ra \sim N^{1/3} \s$ gets further compacted with cross-linker density as $[(u/v)(\phi_c - \phi_c^\ast)/\phi_c^\ast]^{-1/6}$. %  
The solution $\r_0=0$ at $\phi_c< \phi_c^\ast$ corresponds to an open chain following Flory scaling $R_g^0 \approx \s N^{3/5}$. 
Allowing for a bilinear coupling between the monomer and the binder density fields leads to 
$\r_0 \sim \phi_0$ suggesting a non-linear decrease in $\la R_g \ra =R_g^0 [1 + N^{4/5} (z \o /u) \phi_c ]^{-1/3}$~(see Appendix-\ref{ap:notrans}).

\subsubsection{Cluster size}
\label{sec:cluster}
An estimate of the increase in the cluster size of the polymer-bound cross-linkers can be obtained by performing linear stability analysis of Eq.(\ref{eq:dyn}) around a uniform state of $\r=\bar \r$ and $\phi = \bar \phi$. 
This analysis is presented in detail in Appendix-\ref{ap:linstable}. It shows that the uniform state gets unstable towards formation of clusters as an effective coupling strength $\chi = u \bar \r/\o \phi_c^\ast$ crosses a threshold value. The mean spatial extension of such clusters is given by 
\bea
\ell_0  = 2\pi \left[ M_\phi \f{\k w}{r u} \f{1}{\left( 1- \f{\phi_c}{\phi_c^\ast}\right) + 3 \f{v}{u} {\bar \r}^2 }  \right]^{1/4}. \nn
\eea
A uniform density of cross-linkers suggests a mean cluster size $\la C_s \ra \sim \ell_0^3$ leading to
\bea
\la C_s \ra= {\cal A} \left [ \left(1- \f{\phi_c}{\phi_c^\ast} \right) + 3 \f{v}{u} {\bar \r}^2 \right]^{-3/4}. 
\label{eq:cs}
\eea
Replacing $\bar \r = (z \o/u) \phi_c$, the dependence $\la C_s \ra= {\cal A} [(1-\phi_c/\phi_c^\ast +{\cal B} \phi_c^2)]^{-3/4}$ reasonably captures the growth 
in mean cluster size with ${\cal A} = 1.9$ and a small enough ${\cal B} = (3 v z^2 \o^2 / u^3)$ such that ${\cal B} \phi_c^2 \ll 1$,
as the coil-globule transition is approached from below~(Fig.\ref{fig_meanRg}($b$)).

\subsubsection{Time scale}
The diverging time-scale observed in simulations can be understood using the following scaling argument based on Eq.(\ref{eq:dyn}).  For this purpose, we use the length scale associated with the unstable mode $\ell_0$.  Eq.(\ref{eq:dyn}) suggests a relaxation time $\t_r \approx (\ell_0^2/M_\r u) (1-\phi_c/\phi_c^\ast)^{-1}$. Using $\g \phi_c^2 \ll 1$ the relation simplifies to
\bea
\t_r \approx \f{4\pi^2}{M_\r u} \left( M_\phi \f{\k w}{r u} \right)^{1/2} \left[ 1 - \f{\phi_c}{\phi_c^\ast} \right]^{-3/2}, 
\label{eq:taur}
\eea
suggesting a divergence of correlation times as $(1-\phi_c/\phi_c^\ast)^{-3/2}$ near the critical point. For finite sized chains, while the time scales do not diverge, they show significant increase near criticality~(Fig.\ref{fig_meanRg}($c$)). Added with a constant background, Eq.\ref{eq:taur} gives a reasonable description of the simulation results. As is shown in Fig.\ref{fig_phicN}($c$) of Appendix-\ref{ap:transition}, the correlation time at criticality increases with chain length with an approximate power law $\sim N^{9/4}$ indicating divergence.

\section{Local morphology}
\label{sec:morpho}
The binder mediated chromosomal compaction is associated with local morphological changes. The cross-linking due to binders may cause loop formation. In chromosomes, formation of such loops are expected to be highly complex, involving polydispersity of loop-sizes. 
The cross-linkers may also form zipper between contiguous polymeric segments. These, in turn, would enhance contact formation, and as a result modify subchain extensions.  In this section, we discuss the change in all of these three aspects along the phase transition described above.

%%%%%%%%%%%
\begin{figure}[t]
\begin{center}
\includegraphics[width=8cm]{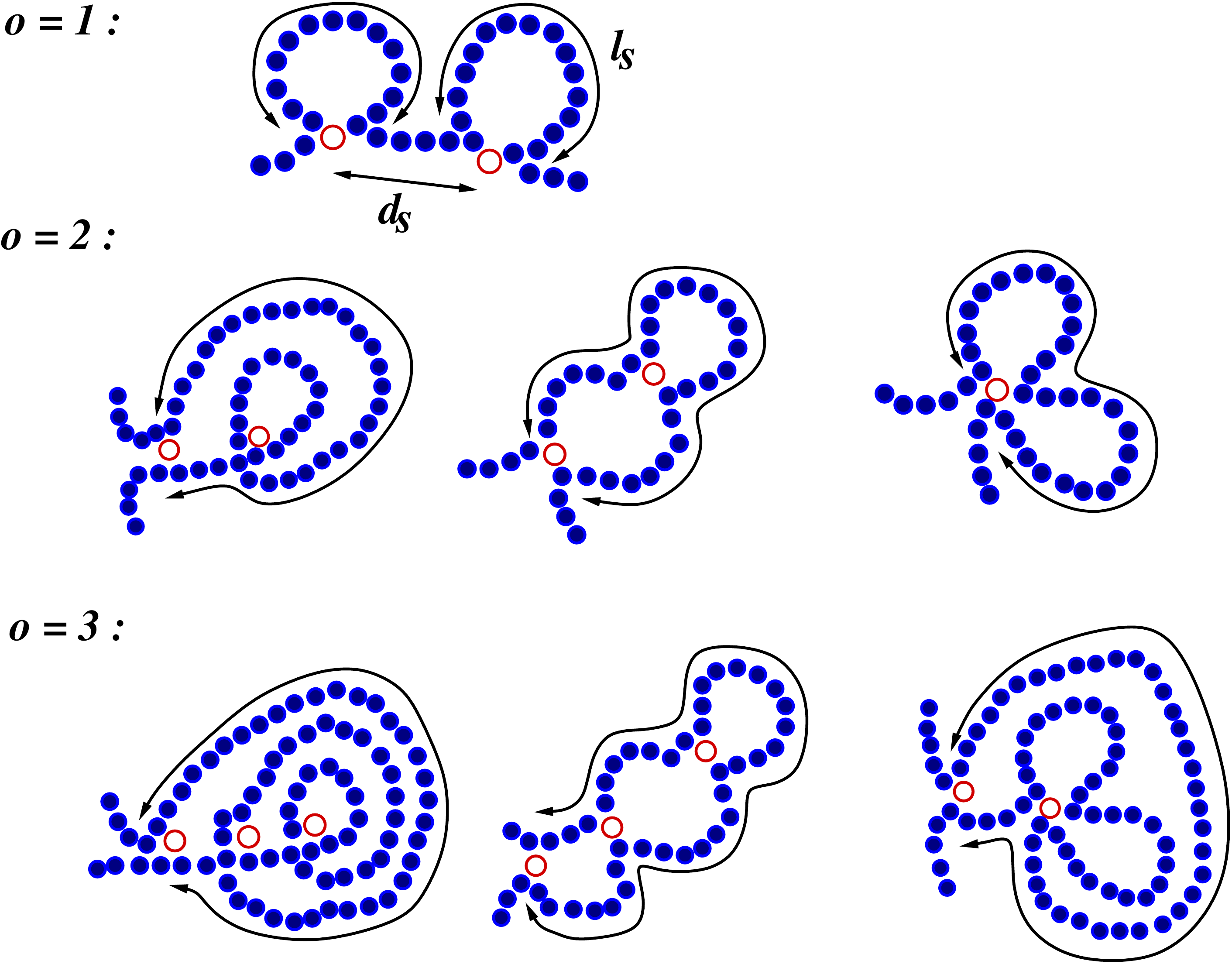}
\caption{(color online) {Schematics of loop topologies of order $o$:} Polymer segments are indicated by blue beads and polymer-bound cross-linkers are shown by red open circles. 
($a$)~Simply connected loops of order $o=1$. Two first order loops of size $l_s$ are separated by a gap of size $d_s$. 
($b$)~Three examples of $o=2$ loops. In the first two cases, the second order loop has one $o=1$ loop embedded inside. The third case shows two embedded  $o=1$ loops.
($c$)~Three examples of $o=3$ loops. In the first two cases, the third order loop has a $o=1$ and a $o=2$ loop embedded. The third case shows two first order loops and a second order loop embedded inside the $o=3$ loop. 
}
\label{fig:cartoon}
\end{center}
\end{figure}
\begin{figure}[h]
\begin{center}
\includegraphics[width=8cm]{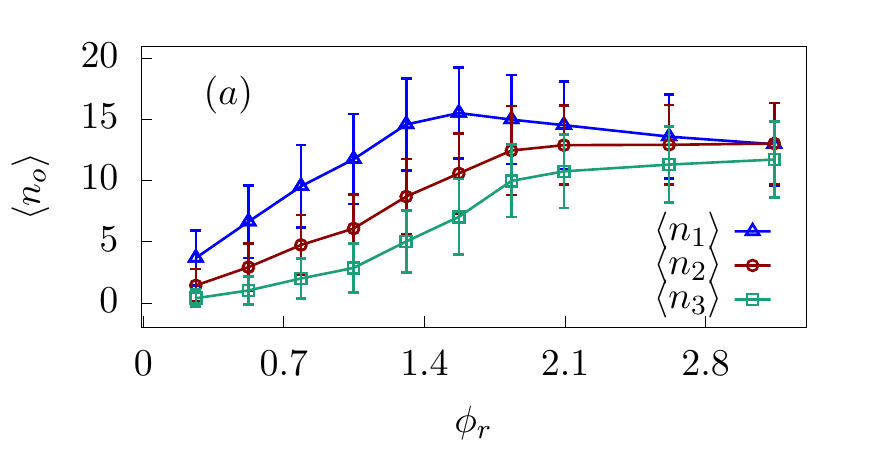}
\includegraphics[width=8cm]{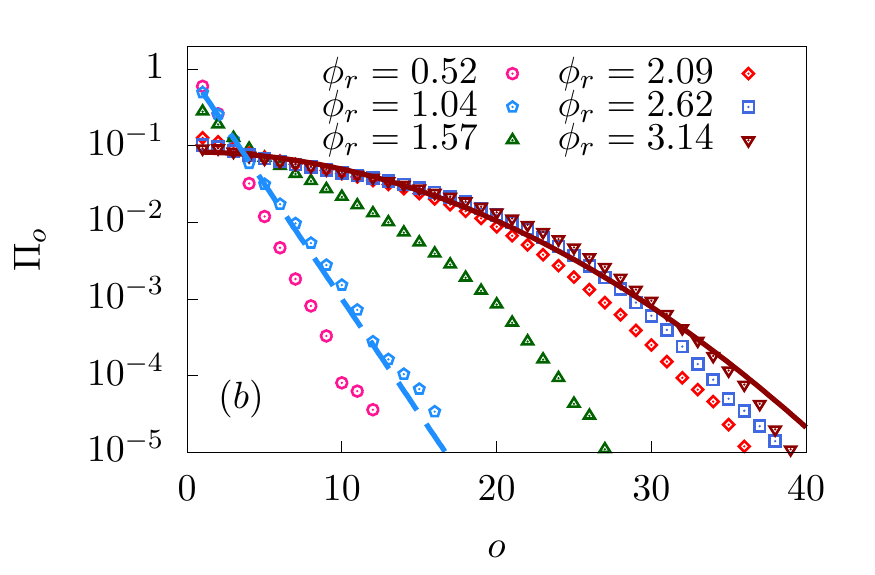}
\caption{(color online) ($a$)\,Mean number of $o$-th order loops $\la n_o\ra$ as a function of density of cross-linkers $\phi_r=\phi_c\times 10^3$.  
Here $\la n_{1,2,3} \ra$ denote the mean number of first, second and third order loops. 
($b$)\,Probability of $o$-th order loop  $\Pi_o$ is plotted on semi-log scale, for various cross-linker densities denoted in the labels. 
At $\phi_r$ = 3.14, the probability of higher order loops decays with an approximate Gaussian form $\exp(- o^2/ 2 g^2)$ where the standard deviation $g=9.83$ (solid brown line). 
For $\phi_r$ = 1.04, the probability of higher order loops decays exponentially as $\exp(-o/\bar o)$ with $\bar o = 1.45$ (dashed blue line). 
}
\label{fig:prob_loop_order}
\end{center}
\end{figure}
%%%%%%%%%%%%%%

\subsection{Loops}
\label{sec:loop}

We describe the possible loop-topologies with the help of Fig.\ref{fig:cartoon}.  A simply connected, or, first order loop is formed by a cross-linker binding two segments of the polymer in such a way that if one moves along the chain from one such segment to the other, no other cross-linker is encountered on the way. With removal of the cross-linker- bond stabilizing such a loop, the first order loop itself disappears (Fig.\ref{fig:cartoon}:~$o=1$). In the figure, $l_s$ and $d_s$ denote loop-size and gap-size between such loops, respectively. In numerical evaluation of mean $d_s$, all intermediate higher order loops are disregarded.  

A higher order loop denoted by order $o=n$, embeds all possible lower order loops $o=1,\dots,(n-1)$ within it. In Fig.\ref{fig:cartoon}:~$o=2$, three examples of second order loops are shown. 
In the first two examples removing one cross-linker reduces the second order loop to a first order loop. In the third example of $o=2$ loop, three bonds of a single cross-linker maintains the loop, and with its removal the whole loop structure disappears.   
In Fig.\ref{fig:cartoon}:~$o=3$ we show three examples of third order loops. 
Note that the first order and higher order loops identified here are related to the serial and parallel topologies described in Ref.~\cite{Mashaghi2014}. 
As it has been shown before, consideration of chromosomal loops is crucial in understanding of its emergent behavior~\cite{Chaudhuri2012, Chaudhuri2018, Swain2019, Wu2018}. 
In this paper, we restrict ourselves to the relative importance of different orders of loops in local chromosomal morphology.

\begin{figure}[t]
\begin{center}
\includegraphics[width=8.6cm] {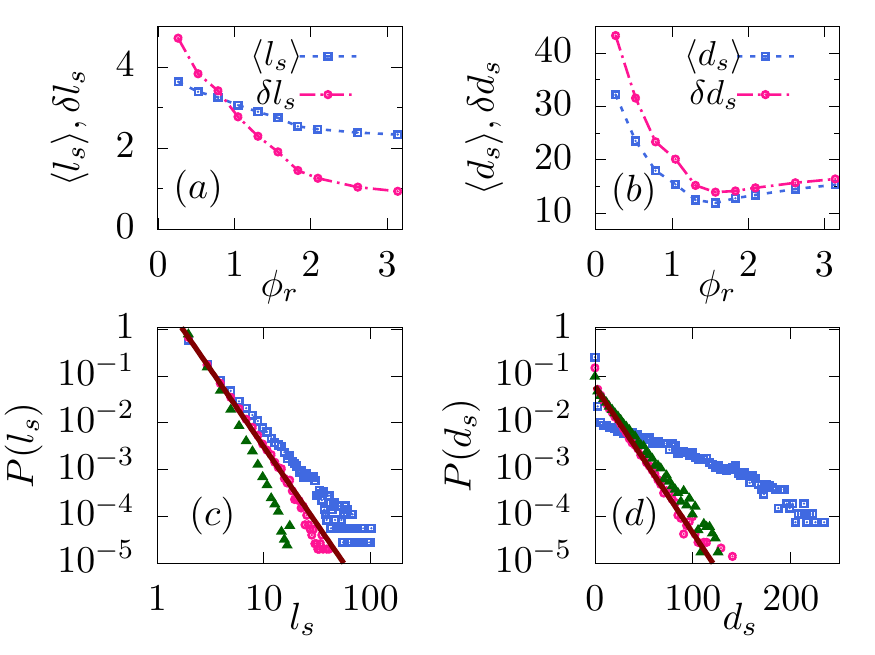}
\caption{(color online) ($a$)~Decrease of the mean size of first order loops $\la l_{s} \ra$ and its fluctuations $\delta l_{s}$ with $\phi_r= \phi_c \times 10^3$. 
($b$)~Non-monotonic variation of mean separation between first order loops $\la d_{s} \ra$ and its fluctuations $\delta d_{s} $ with $\phi_r$. 
Probability distributions of the size of first order loops $P(l_{s})$ and gaps between them $P(d_{s})$ are plotted in ($c$) and ($d$) at  $\phi_r = 0.26\,({\color{blue}\Box}),\, 1.57\, ({\color{magenta}\circ}), \pi\, ({\color{green}\triangle})$. 
At the transition point $\phi_r = 1.57$, $P(l_s) \sim l_s^{-3.3}$,  and
$P(d_s) \sim \exp(-d_s/\l)$ with $\l = 13.8\, \s$, shown by the solid (brown) lines in ($c$) and ($d$) respectively. }
\label{fig:loop_sz}
\end{center}
\end{figure}
   
In Fig.\ref{fig:prob_loop_order}($a$) the mean number of loops $\la n_o \ra$ of order $o=1,2,3$ are shown against the cross-linker density $\phi_c$. All through, $\la n_1 \ra$ remains larger than $\la n_{o=2,3} \ra$ corresponding to higher order loops that show a sigmoidal dependence on $\phi_c$. Interestingly,  $\la n_1 \ra$ maximizes at the phase transition point $\phi_c^\ast$. Thus at the critical point the local morphology of the model chromosome is dominated by the first order loops.

Fig.\ref{fig:prob_loop_order}($b$) shows the probability $\Pi_o$ of a loop to be of $o$-th order. At small cross-linker densities $\phi_c < \phi_c^\ast$, the probability of higher order loops fall exponentially as $\Pi_o = \exp(-o/\bar o)$.  This behavior changes qualitatively after the coil-globule transition ($\phi_r=1.57$) to a Gaussian profile $\exp(-o^2/2 g^2)$, as is shown in Fig.\ref{fig:prob_loop_order}($b$) . 

Given that loop sizes could be measured from electron microscopy~\cite{Postow2004}, we further analyze the statistics of loop-sizes and inter-loop gaps corresponding to the first order loops in Fig.~\ref{fig:loop_sz}. 
With increasing cross-linker density $\phi_c$, the mean size of first order loops $\la l_s \ra$ decreases~(Fig.~\ref{fig:loop_sz}($a$)), as their number increases~(Fig.\ref{fig:prob_loop_order}($a$)) reducing the mean gap size $\la d_s \ra$~(Fig.\ref{fig:loop_sz}($b$)). However, increased $\phi_c$ stabilizes the loops, shown by decreased fluctuation of loop-sizes $\d l_s = \sqrt{\la l_s^2\ra - \la l_s\ra^2}$. The mean gap size $\la d_s \ra$ and its fluctuation $\d d_s = \sqrt{\la d_s^2\ra - \la d_s\ra^2}$ reach their minimum at the transition point $\phi_c^\ast = 1.57 \times 10^{-3}$. The increase in the inter-loop separation $\la d_s \ra$ beyond this point is due to the increase in probability of higher order loops in the local morphology of the model chromatin. 

Fig.\ref{fig:loop_sz}($c$) and ($d$) show the probability distributions of first order loop sizes $P(l_{s})$, and separation between  consecutive first order loops $P(d_{s})$, respectively. For all $\phi_c$ values, $P(l_s) \sim l_s^{-\mu}$, with $\mu$ increasing with $\phi_c$ in a sigmoidal fashion, giving $\mu=3.3$ at the critical point $\phi_c^\ast=1.57 \times 10^{-3}$. The power law distribution of $P(l_s)$ shows that their is no characteristic loop size, and loops of all possible lengths are present. On the other hand,  the gap size distributions follow an approximate exponential form $P(d_s) \approx (1/\la d_s\ra) \exp(-d_s/\la d_s \ra)$.

\label{sec:loop}
\begin{figure}[h!]
\begin{center}
\includegraphics[width=8cm] {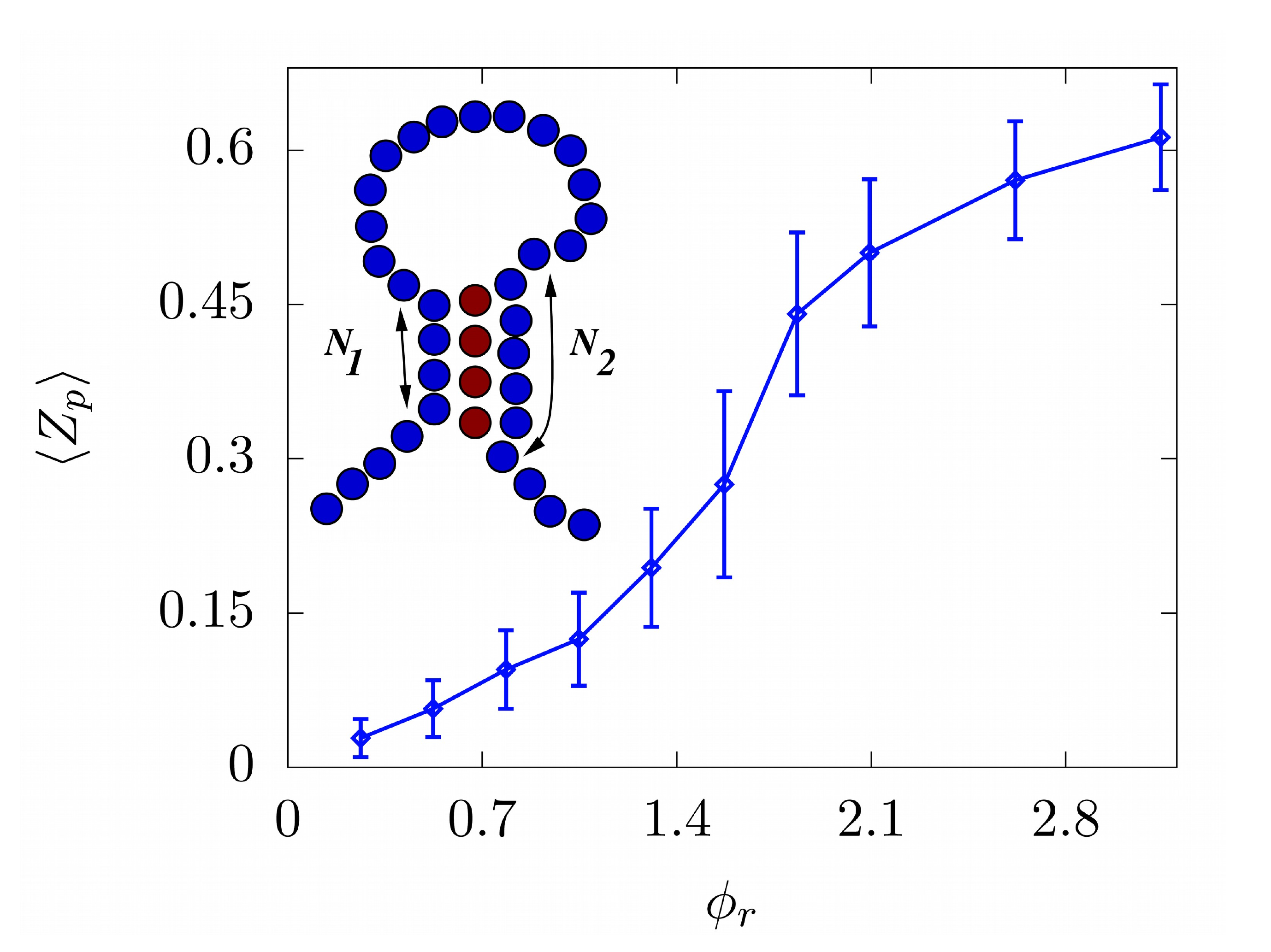}
\caption{(color online) The zippered fraction of model chromosome $\la Z_p \ra$ as a function of the cross-linker density $\phi_r = \phi_c \times 10^3$. The inset shows two contiguous segments containing $N_1$ and $N_2$ monomers (blue beads) forming a zipper via binders (red beads). The corresponding zipper fraction is $Z_p = (N_1+N_2)/N$.  }
\label{fig:zipper}
\end{center}
\end{figure}
\subsection{Zippering}
The binders can also zipper different segments of the polymer. The inset of Fig.\ref{fig:zipper} shows one such zipper maintained by cross-linkers. The zipper fraction of a conformation is given by $Z_p = (1/N) \sum_{\xi,i} N_i^\xi$, where  $N_i^\xi$ are the number of monomers involved in forming $\xi$-th zipper, and $N$ is the total number of monomers in the chain.  Fig.\ref{fig:zipper} shows variation of ensemble averaged zippered fraction with the cross-linker density. 
The zipper fraction increases non-linearly to saturate in the equilibrium globule phase to a value that remains within $60\%$ of the completely zippered filament  $\la Z_p \ra=1$. Near the critical point of the coil-globule transition $\la Z_p \ra \approx 0.3$, half the saturation value. 

\subsection{Contacts and subchain extensions}
Conformational relaxation of a polymer brings contour wise distant parts of the chromatin in contact with each other, even in absence of binders. An example of such a contact in our model was shown in Fig.\ref{fig:snapshot}($c$).  The processes of cross-linking and zippering increase the probability of contact formation that shows an asymptotic behavior $\Pi_c(s)\sim s^{-\a}$ between segments separated by a contour length $s$.    With increasing cross-linker density $\phi_c$, the exponent $\a$ decreases non-linearly to vanish, capturing the asymptotic plateauing of $\Pi_c(s)$ at large $\phi$~(Appendix-\ref{ap:contact}). At criticality, $\a \approx 1.1$, similar to the fractal globule and human chromosomes~\cite{Lieberman-Aiden2009, Mirny2011, Grosberg1988}. This behavior is in agreement with an earlier lattice model~\cite{Barbieri2012}.  The mean subchain extension shows asymptotic power law $\la r^2(s) \ra \sim s^{2\nu}$. The exponent $\nu$ reduces from $3/5$ at $\phi_c=0$ to the fractal-globule like $\nu \approx 1/3$ at the critical point. At the highest cross-linker densities, $\nu = 0$, a behavior typical of equilibrium globules~(Appendix-\ref{ap:subchain}).  Finally, the structure of the detailed contact-map changes with $\phi$~(Appendix-\ref{ap:contact_map}). At  small $\phi_c$, only contour-wise neighbors participate in contact formation. However, at the transition point $\phi_c = \phi_c^\ast$, contacts begin to percolate to monomers separated by long contour lengths.

\section{Discussion}
\label{sec:concl}
In summary, using an off-lattice model of self avoiding polymer and diffusing protein binders cross-linking different segments of the chromatin fibre, 
we have presented an extensive characterization of the continuous chromatin folding transition, and analyzed the associated changes in chromatin morphology in terms of formation of  loops, zippering and contacts.  
The criticality is characterized by unimodal distributions, divergent fluctuations and critical slowing down. 
The negative maximum in the cross-correlation between the number of attached binders and chromosome size, at criticality, might be 
utilized by living cells for easy switching between folded and open conformations, providing easy access to DNA-tracking enzymes. This is suggestive of a possibility that chromosomes might be poised at criticality~\cite{Mora2011}, vindicated further by the similarity of the calculated contact probability at the critical point with the average behavior of human chromosomes. 
Although the local chromatin morphology does show highly complex loop structures, at criticality,  it is dominated by simply connected loops. 

Each coarse-grained chromatin bead in our model can be considered as $10 - 12$ closely packed nucleosomes containing around $2 - 2.5\,$kbp DNA-segments having a diameter $\s\approx 20 - 40\,$nm~\cite{Routh2008, Mirny2011}.  
The dimensionless critical volume fraction $\phi_c$ is equivalent to a concentration $[\phi_c^\ast/(4 \pi \s^3/3)]$,  
which can be expressed in terms of molarity by dividing it by the Avogadro number. This leads to the estimate of  critical concentration between $\sim  60\,$nmol/l\,$ - 470\,$nmol/l. 
The mean size of the first order loops observed at criticality translates to $4 - 7\,$kbp. 
The estimated ratio of this loop size and inter-loop gaps is $\la l_s \ra: \la d_s \ra \approx  1:5$ at this concentration. 

{ In the chromosomal environment having viscosity $\eta$, the dissipation constant $\g=3 \pi \eta \s$. 
As it has been observed, the nucleoplasm viscosity $\eta$ felt by objects within the cell nucleus depends on their size~\cite{Lukacs2000,Tseng2004}. }
Using the measured viscosity $\sim 10\,$Pa-s felt by solutes having $\sim 10\,$nm size~\cite{Tseng2004} for the $\s=20\,$nm beads, the { characteristic time which is the same as the time required to diffuse over the length-scale $\s$ can be determined by using the relation}  $\t=\g \s^2/\kb T = 0.2\,$s. 
Thus,  the simulated correlation time $\t_{R_g}$ denoting chromosomal relaxation over $\sim 0.5 - 0.6\,$Mbp  translates to $\approx 22\,$minutes at the critical point.

Here we should reemphasize that our study represents an average description of chromosomes using a  coarse grained homopolymer model. This approach did not aim to distinguish interaction between specific protein types and gene sequences. 
While some of our predictions appear to compare well with experiments, others
involving cross-linker clusters, relaxation time, and loop morphology are amenable to  experimental verifications.  

 \acknowledgments
The simulations were performed on SAMKHYA, the high performance computing facility at IOP, Bhubaneswar. 
 DC acknowledges SERB, India, for financial support through grant number EMR/2016/001454, ICTS-TIFR, Bangalore for an associateship, and thanks Bela M. Mulder for useful discussions. 
 We thank Abhishek Chaudhuri for a critical reading of the manuscript.

\appendix

\section{Valency of cross-linkers}
\label{sec:valency}
\begin{figure}[h]
\begin{center}
\includegraphics[width=7cm]{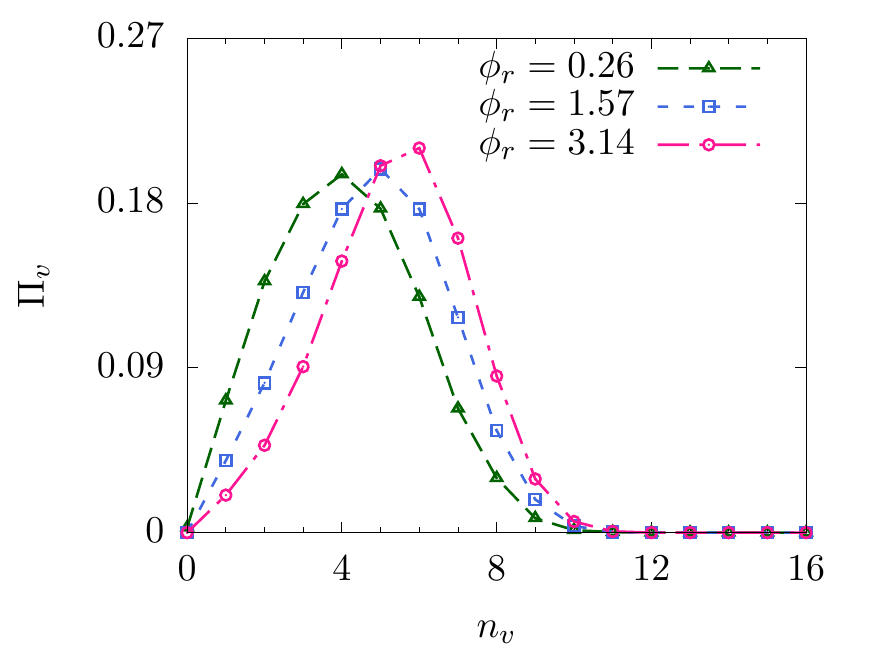}
\caption{(color online) $\Pi_{v}$ represents probability that a cross-linker is simultaneously attached to $n_v$ number of monomers, at cross-linker densities $\phi_r = \phi_c \times 10^3$ denoted in the figure. The maximum shifts from $n_v=4$ to $6$ across the coil-globule transition. At the transition point, $\phi_r=1.57$, the maximum probability corresponds to the valency $n_v=5$.}
\label{fig:valency}
\end{center}
\end{figure}

The cross-linkers used in the simulations are potentially multivalent. Here the question we ask is how many monomers of the chain, a cross-linker attaches to simultaneously? From the simulations, we identify the polymer bound cross-linkers, count the number of monomers that lie within the range of attraction $r_c=1.5\,\s$ identifying the instantaneous valency of a cross-linker, and compute the histogram over all the cross-linkers and time. This leads to the probability $\Pi_v$ of valency $v$, normalized to $\sum_v \Pi_v=1$.  The maximum of the probability indicates the typical valency of cross-linkers at an ambient density $\phi_c$~(see Fig.\ref{fig:valency}). As  $\phi_c$ increases, the peak shifts towards larger values. It means with increase of $\phi_c$, a single cross-linker on an average binds to more number of monomers of the chain. In the fully compact state, at the largest $\phi_c$, the typical valency we find is $6$.

\begin{figure}[t]
\begin{center}
\includegraphics[width=8cm]{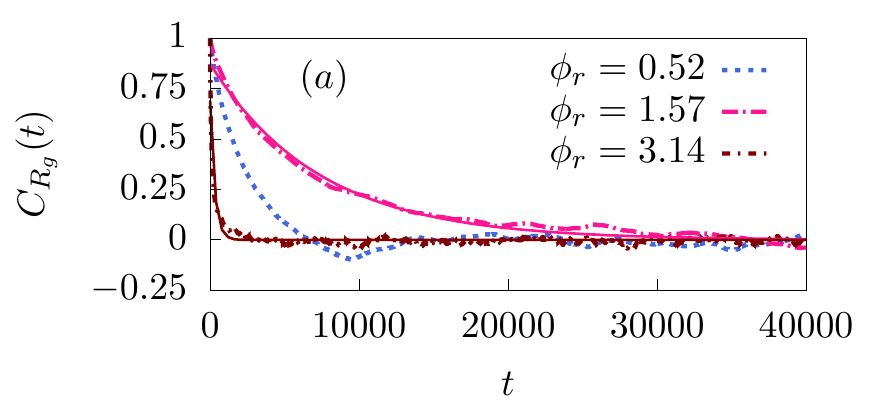}
\includegraphics[width=8cm]{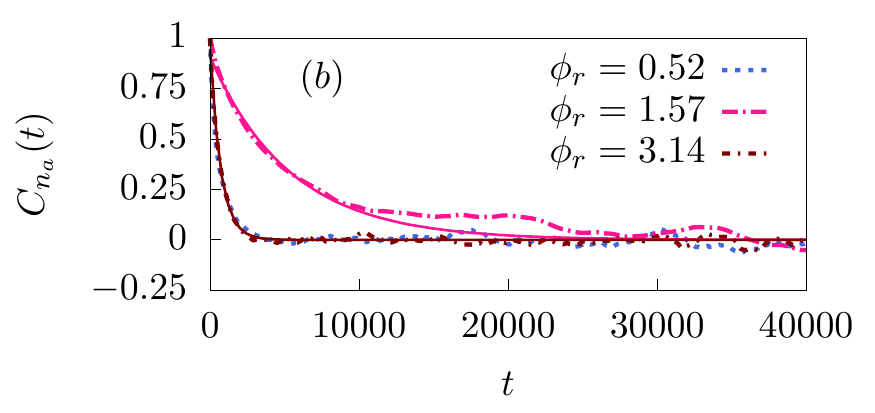}
\caption{(color online) The auto-correlation functions   ($a$)\,$C_{R_g}(t)$ of polymer radius of gyration $R_g$, and ($b$)\,$C_{n_a}(t)$  of the number of  cross-linkers attached to the chain $n_{a}$, at three cross-linker densities $\phi_r = \phi_c \times 10^3$; $t$ is expressed in unit of $\t$. Fitting them to exponential forms $\exp(-t/\t_c)$ gives correlation time $\t_c = \t_{R_g},\, \t_{n_a}$ for $R_g$ and $n_a$ respectively. 
Two such fittings are shown in each plot by solid lines. 
The fitted correlation times are  $\t_{R_g}=7371\, \t$, $\t_{n_a}=5402\, \t$ at $\phi_r=1.57$, and $\t_{R_g}=307\, \t$, $\t_{n_a}=720\,\t$ at  $\phi_r=\pi$.
}
\label{fig:correlation}
\end{center}
\end{figure}
%%%%%%%%%%%
\begin{figure}[t]
\begin{center}
\includegraphics[width=8cm]{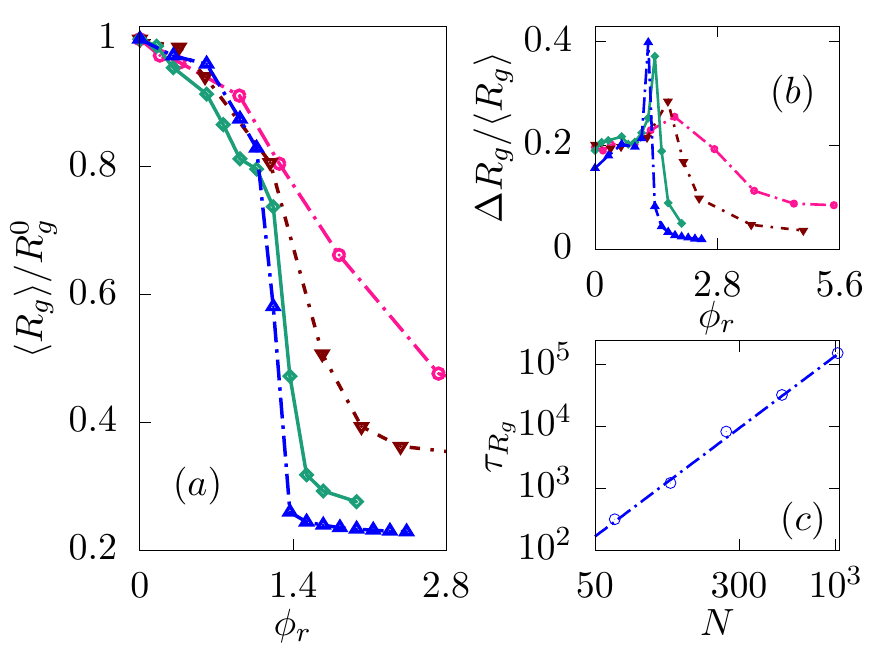}
\caption{(color online) Variation of ($a$)~the scaled radius of gyration $\la R_g \ra/R_g^0$, and ($b$)~its relative fluctuations $\D R_g/ \la R_g \ra$  with cross-linker density $\phi_r= \phi_c \times 10^3$ for chain lengths $N\,= \, 128\, ({\color{magenta}\bigcirc}), \, \, 256\, ({\color{brown}\bigtriangledown}),\, 512\, ({\color{green} \diamond}), 1024\, ({\color{blue}\bigtriangleup})$. ($c$)~The correlation time at the transition point $\t_{R_g}$ increases with chain length $N$.  The dash-dotted line denotes $\t_{R_g} \sim N^{9/4}$.
}
\label{fig_phicN}
\end{center}
\end{figure} 
%%%%%

\section{Correlation function}
\label{sec:corrln}
In Fig.\ref{fig:correlation}, we present normalized auto-correlation functions of the polymer radius of gyration $C_{R_g}(t)= \la \d R_g(t) \d R_g(0)\ra/\la \d R_g^2 \ra$, and the total number of bound cross-linkers $C_{n_a}(t)=\la \d n_a(t) \d n_a(0) \ra/\la \d n_a^2 \ra$  at various cross-linker densities $\phi_c$ across the coil-globule transition. The correlations show approximate exponential decay $\exp(-t/\t_c)$ with correlation time $\t_c$ denoted by $\t_{R_g}$ for the polymer radius of gyration, and $\t_{n_a}$ for the total number of polymer bound cross-linkers. The fitted values show a maximum at the phase transition point $\phi_c^\ast$ as is shown in Fig.~\ref{fig_meanRg}($c$).

\section{System size dependence at the coil-globule transition}
\label{ap:transition}

We performed simulations with various chain lengths $N$ to check the system size dependence on the coil-globule transition. We observe a continuous change of $\la R_g \ra$ with cross-linker density $\phi_c = \phi_r \times 10^{-3}$~(Fig.~\ref{fig_phicN}($a$)\,) with the transition becoming sharper at larger $N$. The relative fluctuations $\D R_g/R_g$ at the transition point increases with $N$~(Fig.\ref{fig_phicN}($b$)\,). The correlation time  $\t_{R_g}$ characterizing fluctuations in $R_g$ at the transition point  diverges with the increase in polymer size $N$. The simulations suggest a dynamical scaling $\t_{R_g} \sim N^{\zeta}$ with $\zeta \approx 9/4$.

\section{Before transition}
\label{ap:notrans}
The most general mean field theory would also contain terms bilinear in $\r$ and $\phi$, which we neglected in the discussion of phase transition. In presence of such a coupling, keeping terms only up to quadratic order, 
\
\bea
\be f =   \hf  u\, \r^2  + \hf w \phi^2 - z \r \phi + \f{\k}{2} (\nabla \r)^2. \nn
\eea
This does not describe any phase transition, however, suggests a uniform mean field solution $ \r_0 = z \phi_0/u = z \o \phi_c/u$.  Thus, before the transition, the mean radius of gyration is expected to decrease with $\phi_c$  as $\la R_g \ra = R_g^0 [1 + N^{4/5} \r_0]^{-1/3} = R_g^0 [1 + N^{4/5} (z \o /u) \phi_c ]^{-1/3}$.

\section{Linear stability analysis}
\label{ap:linstable}
The formation of cross-linker clusters mediates folding of the model chromosome.  
To characterize the dynamics, we use linear stability analysis for small deviations from a homogeneous state $\r = \bar \r+\d \r(\rv)$, $\phi = \bar \phi + \d \phi(\rv)$. The dynamics in Eq.(\ref{eq:dyn}) for these small deviations become
\bea
\p_t \d\r &=& D_\r \nabla^2 \d \r - M_\r \k \, \nabla^4 \d \r - M_\r \chi \, \nabla^2 \d \phi \nn\\
\p_t \d\phi &=& D_\phi \, \nabla^2  \d \phi - M_\phi \chi\,  \nabla^2 \d \r - r\, \d\phi, \nn
\eea
where $$D_\r = M_\r u \left[ \left( 1- \f{\bar \phi}{\phi_\ast}\right) + 3 \f{v}{u} \bar \r^2\right], $$  and $D_\phi=M_\phi w$ 
are the effective diffusion constants of the two components, and $\chi=u\bar \r/\phi_\ast$ is the strength of cross-coupling. In the above equations $\p_t$ denotes the partial derivative with respect to time $t$. 
Expressing time in units of inverse turnover rate, $\t_u = 1/r$, and lengths in units of $x_u = \sqrt{M_\phi w/ r}$, one finds
\bea
\p_\t \d\r &=& {\cal D}_0 \nabla_\xi^2 \d \r - {\cal K} \nabla_\xi^4 \d \r - {\cal C} \nabla_\xi^2 \d\phi \nn\\
\p_\t \d \phi &=& \nabla_\xi^2  \d \phi - {\cal C}' \nabla_\xi^2  \d \r - \d \phi,
\label{eq:lin2}
\eea
with control parameters of the dynamics ${\cal D}_0 = D_\r/M_\phi w$, 
${\cal K} = \f{M_\r}{ M_\phi^2} \, \f{\k r}{w^2} $, 
${\cal C}=\f{M_\r}{ M_\phi} \f{\chi}{w} $, and
${\cal C}' = \f{\chi}{w} $. The dimensionless time and length scales are denoted by $\t = t/\t_u$, and $\xi = x/x_u$, respectively. 

Fourier transform of this equation gives evolution of modes as matrix equations $\p_\t\, (\d \r_q, \d \phi_q) = {\cal M} \, (\d \r_q, \d \phi_q)$, where,
\bea
{\cal M} = 
\begin{pmatrix} 
-q^2({\cal D}_0 + {\cal K} q^2) &  {\cal C} q^2 \\
{\cal C}' q^2 & -(q^2+1)\nn
\end{pmatrix}.
\eea
The eigenvalues of ${\cal M}$ are given by 
\bea
\l(q^2) = \hf \left\{ {\rm Tr}{\cal M} \pm \sqrt{ ({\rm Tr}{\cal M})^2 - 4\, {\rm det}{\cal M} } \right\} \nn
\eea
As the trace of this matrix 
$$ {\rm Tr.} {\cal M} = -q^2({\cal D}_0 + {\cal K} q^2)-(q^2+1)  \, < \, 0, $$ 
the only way of having instability (one of the eigenvalues becomes positive) is if the determinant 
$$ {\rm det} {\cal M} = q^2(q^2+1)({\cal D}_0 + {\cal K} q^2) - {\cal C} {\cal C}' q^4 <0.$$ 
This last criterion leads to ${\cal C} {\cal C}' > F(q^2)$ where $F(q^2)=(1+\f{1}{q^2})({\cal D}_0 + {\cal K} q^2)$. This will be satisfied for any $q^2$ if even the minimum of $F(q^2)$ 
obeys this inequality. One can easily show that $F(q^2)$ is minimized at $q_0^2 = \sqrt{{\cal D}_0/{\cal K}}$, and $F(q_0^2) = (\sqrt{{\cal D}_0} + \sqrt{{\cal K}})^2$.  Thus the instability criterion becomes
\bea
\sqrt{ {\cal C} {\cal C}'} > (\sqrt{{\cal D}_0} + \sqrt{{\cal K}}). \nn
\eea
Note from Eq.(\ref{eq:lin2}) that ${\cal C}$ and ${\cal C}'$ denote coupling coefficients between the evolution of the two fields $\r$ and $\phi$. 
The inequality suggests a minimal coupling strength $\chi$ is required to generate instability towards  formation of cross-linker clusters,
\bea
\chi > \sqrt{\f{\k r}{M_\phi}}  + \sqrt{u w \left[ \left(1-\f{\bar \phi}{\phi_\ast}\right) + 3 \f{v}{u} \bar \r^2  \right]}.  \nn
\eea

Once this condition is satisfied, instability in the form of clustering of cross-linkers, mediated by the attractive interaction with monomers, arise. The fastest growing mode $q_0 = ({\cal D}_0/{\cal K})^{1/4}$ 
 predicts the most unstable length scale $\ell_0/x_u = 2\pi/q_0 = 2\pi ({\cal K}/{\cal D}_0)^{1/4}$, which gives the mean extension of the clusters %size in physical units
\bea
\ell_0  = 2\pi \left[ M_\phi \f{\k w}{r u} \f{1}{\left( 1- \f{\bar \phi}{\phi_\ast}\right) + 3 \f{v}{u} \bar \r^2 }  \right]^{1/4}. \nn
\eea

%%%%%%%%%
\begin{figure}[t]
\begin{center}
\includegraphics[width=8cm]{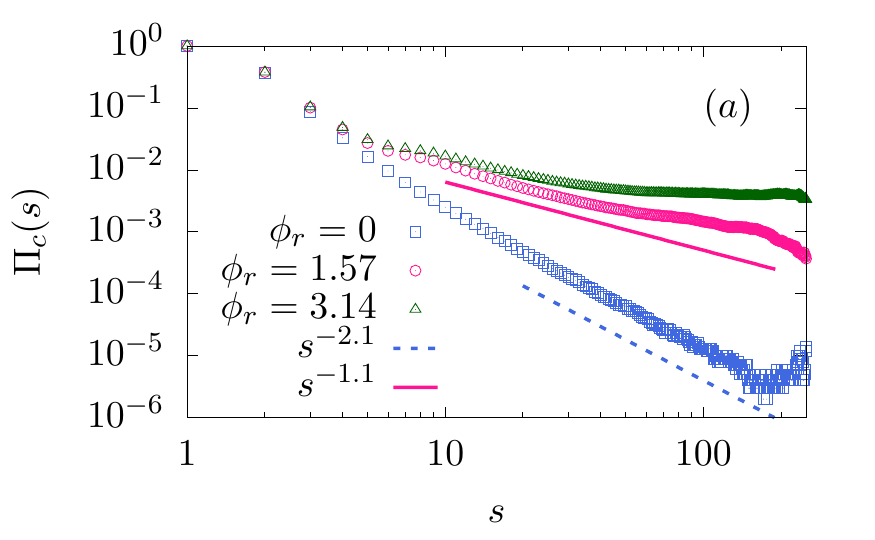}
\includegraphics[width=8cm]{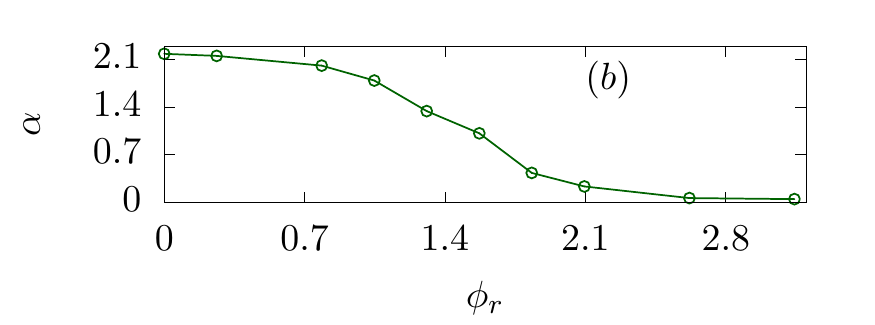}
\caption{(color online) ($a$)~Contact probabilities $\Pi_c(s)$ at different cross-linker densities  $\phi_r = \phi_c \times 10^3$. They follow asymptotic power law profiles $\Pi_c(s) \sim s^{-\a}$ at all $\phi_c$, with $\a$ being a function of $\phi_c$. 
($b$)~The decrease of exponent $\a$ with increasing $\phi_r$ is related to the coil-globule transition.}
\label{fig:contact_sz_distbn_expo}
\end{center}
\end{figure}
%%%%%%%%%
\section{Contact probability} 
\label{ap:contact}

To analyze contact formation from simulations one requires a finite cutoff length such that if two monomers fall within such a separation they are defined to be in contact. Here we use 
$r_c=1.5\,\s$ for this purpose.
We have checked that our main results do not depend on the precise choice of this length scale.
 
The contour wise separation between two monomers, $s$, defines the genomic distance between chromatin segments. 
The contact probability $\Pi_c(s)$ is a measure of such two segments to be in contact. 
In absence of binders, we get $\Pi_c(s) \sim s^{-\a}$ with $\a \approx 2.1$, as expected for self-avoiding chains~\cite{Gennes1979}.
Even in presence of cross-linkers, the asymptotic power law persists with $\phi_c$-dependent $\a$~(Fig.\ref{fig:contact_sz_distbn_expo}.($a$)). 
At the transition point $\phi_c^\ast = 1.57 \times 10^{-3}$, the simulation results are consistent with $\a \approx 1.1$, a number that agrees well with the prediction of the fractal globule model~\cite{Mirny2011, Grosberg1988}. 
It is interesting to note that $\a \approx 1.1$ is close to the average exponent found across all human cell chromosomes, % $\a \approx 1.08$, 
in the genomic distances of 0.5-10 Mbp range~\cite{Lieberman-Aiden2009,Mirny2011}, and belongs to the range of exponents observed in individual mammalian chromosomes~\cite{Lieberman-Aiden2009, Dixon2012, Barbieri2012}. 
At large $\phi_c$ values, after the completion of the coil-globule transition, contact probabilities at large $s$ plateaus to a constant, indicating $\a = 0$.
As a function of  $\phi_c$, the asymptotic exponent $\a$ reveals a continuous decrease~(see Fig.\ref{fig:contact_sz_distbn_expo}($b$)\,), capturing the change in polymeric organization in the course of the coil-globule transition.

%%%%
\begin{figure}[t]
\begin{center}
\includegraphics[width=8cm]{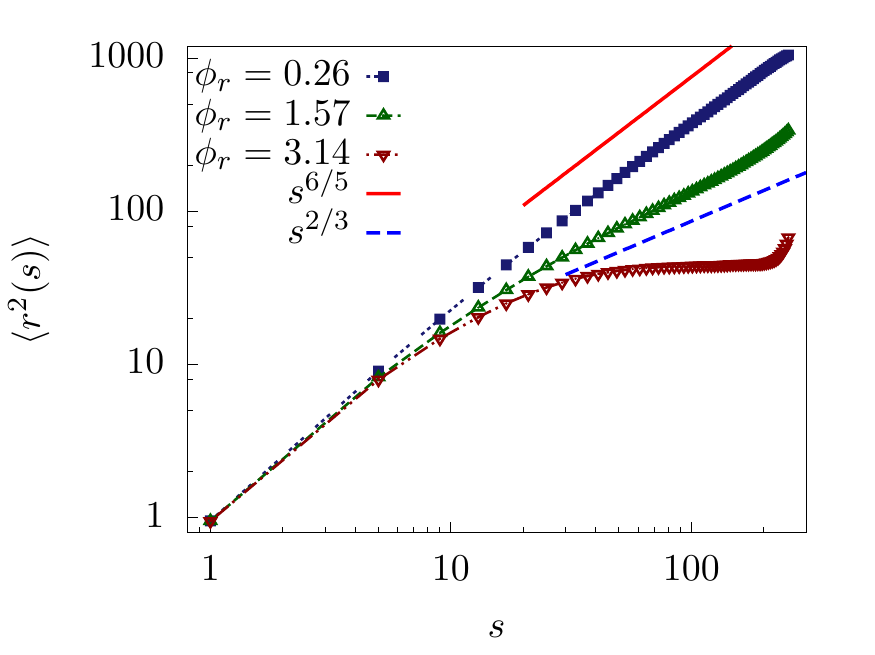}
\caption{(color online) The scaling behavior of subchain extension $\la r^2(s) \ra$, at three cross-linker concentrations $\phi_r = \phi_c \times 10^3$, before, at and after the coil-globule transition. At low densities, it approximately follows Flory scaling $\la r^2(s) \ra \sim s^{6/5}$~(red solid line).  At the transition point $\phi_c = 1.57 \times 10^{-3}$, the asymptotic behavior  agrees with the fractal globule estimate $\la r^2(s) \ra \sim s^{2/3}$~(blue dashed line).  At the highest concentrations we find asymptotic plateauing, a characteristic of equilibrium globules. 
}
\label{fig:rs_scaling}
\end{center}
\end{figure}
%%%%%%%%%%%
\begin{figure*}[t]
\begin{center}
\includegraphics[width=16cm]{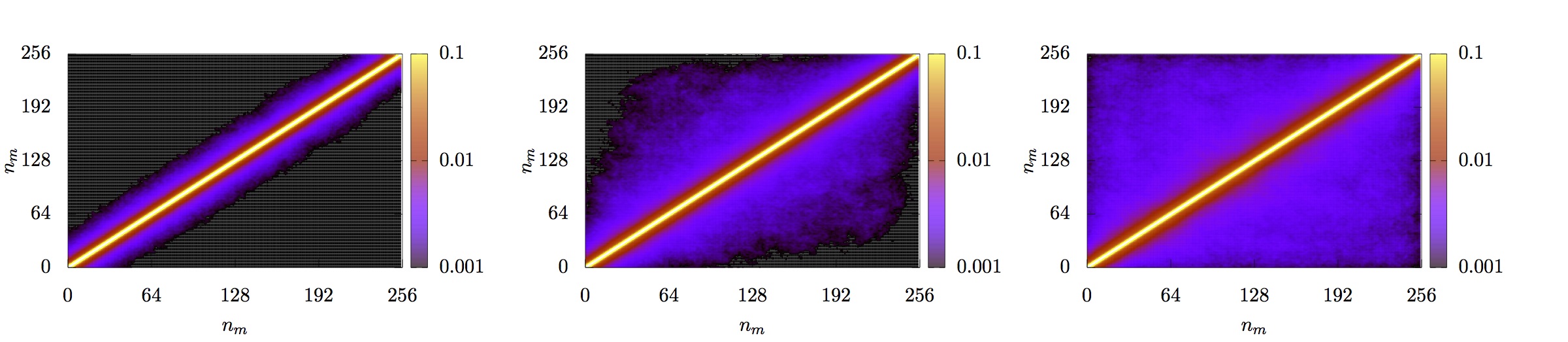}
\caption{(color online) 
From left to right, contact maps at $\phi_c = 1.31 \times 10^{-3}, 1.57 \times 10^{-3}$, and $1.83 \times 10^{-3}$ are plotted. The color code captures contact frequency and is shown in log scale.
}
\label{fig:contact_map}
\end{center}
\end{figure*}
%%%%%%%%%%

\section{Extension of subchains}
\label{ap:subchain}
%%%%%%%%%%
Here we consider the scaling behavior of subchain extensions, measured in terms of the mean squared end to end distance $\la r^2(s) \ra$ in subchains of contour length $s$. We observe three different scaling behaviors across the coil-globule transition. 

A sub-chain inside a compact equilibrium globule is expected to behave like a random walk due to strong screening of interaction by large monomeric density. Thus $\la r^2(s) \ra \sim s$, before the globule boundary is encountered. Multiple reflections from the globule boundary, as $s > \la r^2(s) \ra \sim N^{2/3}$, fills the space inside the globule uniformly, so that it becomes equally likely to find the other end of the subchain anywhere inside the globule, saturating $\la r^2(s)\ra$ to a constant. Thus in equilibrium globules $\la r^2(s) \ra \sim s$ up to $s < N^{2/3}$, and saturates beyond that length scale~\cite{Gennes1979, Mirny2011}. 
The random loop model, with fixed probability of attraction between monomers,  shows all the features of equilibrium globule in final configurations~\cite{Bohn2007,Mateos-Langerak2009}.
On the other hand, the fractal globule is space filling at all scales, such that $\la r^2(s) \ra \sim s^{2/3}$~\cite{Grosberg1988,Mirny2011}.

At small $\phi_c$~($=0.26 \times 10^{-3}$), we find  
a behavior typical of open chains, $\la r^2(s) \ra \sim s^{6/5}$, that follows Flory scaling.  
In the fully folded compact phase at high $\phi_c$~($=\pi \times 10^{-3}$), $\la r^2(s) \ra$ shows plateauing at large $s$ as in compact equilibrium globules, and random loop models~\cite{Mirny2011,DeGennes1975,Bohn2007,Mateos-Langerak2009}. 
Such plateauing  was earlier related to folding of chromosome into territories~\cite{Cremer2001}.
In the compact phase, the molecular cross-linkers may not only pull different segments close to each other, by doing so, they may displace well separated parts further away from each other~\cite{Nicodemi2009}, reflected in the eventual increase of $\la r^2(s) \ra$ as $s$ approaches the full length $N$, e.g., at highest $\phi_c$. 
At the critical point, $\phi_c^\ast$~($=1.57\times 10^{-3}$), simulation results for subchain extensions is consistent with $\la r^2(s) \ra \sim s^{2/3}$ as in fractal globules~\cite{Grosberg1988}.  This is close to  the threshold-exponent predicted in \cite{Barbieri2012} $\la r^2(s) \ra \sim s^{2 \nu}$ with $ \nu =0.39$.
Thus with increasing cross-linker density, the model chromatin morphology changes from an open chain to compact equilibrium globule, via an intermediate fractal globule behavior observed at the critical point. The sustenance of fractal globule like non-equilibrium behavior at the critical point can be understood in terms of the super-slow relaxation.

\section{Contact maps}
\label{ap:contact_map}
In Fig.\ref{fig:contact_map} we present ensemble averaged contact maps over  equilibrium configurations at different cross-linker densities. Such maps represent probability measures of two chromatin segments to be in spatial proximity. At the coil-globule transition $\phi_c = \phi_c^\ast$, the contact shows emergence of local pattern, indicating enhanced probability of contour-wise well separated segments to come into spatial proximity. In the compact phase at $\phi_c = 1.83 \times 10^{-3}$, the chromosomal contacts spread over the whole chromatin chain.

\bibliographystyle{prsty}

%\bibliography{check_1}
\end{document}